# ADAPTATIVE SMOOTH PARTICLE HYDRODYNAMICS AND PARTICLE-PARTICLE COUPLED CODES: ENERGY AND ENTROPY CONSERVATION

A. SERNA

*Laboratoire d'Astrophysique Extragalactique et de Cosmologie, CNRS URA 173, Observatoire de Paris-Meudon, 92195 Meudon, France*
*Service P.T.N., Commisariat à l'Energie Atomique (CEA), Bruyères le Chatel, France*

J.-M. ALIMI

*Laboratoire d'Astrophysique Extragalactique et de Cosmologie, CNRS URA 173, Observatoire de Paris-Meudon, 92195 Meudon, France*

and

J.-P. CHIEZE

*Service P.T.N., Commisariat à l'Energie Atomique (CEA), Bruyères le Chatel, France*

We present and test a general-purpose code, called PPASPH, for evolving self-gravitating fluids in astrophysics, both with and without a collisionless component. In PPASPH, hydrodynamical properties are computed by using the SPH (Smoothed Particle Hydrodynamics) method while, unlike most previous implementations of SPH, gravitational forces are computed by a PP (Particle-Particle) approach. Other important features of this code are: a) PPASPH takes into account the contributions of all particles to the gravitational and hydrodynamical forces on any other particle. This results in a better energy conservation; b) Smoothing lengths are updated by an iterative procedure which ensures an exactly constant number of neighbors around each gas particle. c) Cooling processes have been implemented in an integrated form which includes a special treatment to avoid a non-physical catastrophic cooling phenomenon. Such a procedure ensures that cooling does not limit the timestep. d) Hydrodynamics equations optionally include the correction terms (hereafter $\nabla h$ terms) appearing when $h(t, \mathbf{r})$ is not constant.

Our code has been implemented by using the data parallel programming model on *The Connection Machine* (CM), which allows for an efficient unification of the SPH and PP methods with costs per time step growing as $\sim N$.

PPASPH has been applied to study the importance of adaptive smoothing correction terms on the entropy conservation. We confirm Hernquist's (1993) interpretation of the entropy violation observed in previous SPH simulations as a result of having neglected these terms. An improvement on the entropy conservation is not found by merely considering larger numbers of particles or different $N_S$ choices. The correct continuum description is only obtained if the $\nabla h$ correction terms are included. Otherwise, the entropy conservation is always rather poor as compared to that found for the total energy.

*Subject headings:* hydrodynamics – numerical methods

## 1. INTRODUCTION

Most hydrodynamical problems require numerical calculations because of their complexity. Several numerical methods have been developed to solve the equations of hydrodynamics, however one of the best suited for astrophysical problems is the Smoothed Particle Hydrodynamics (SPH) technique.

SPH is an N-body integration scheme introduced by Lucy (1977) and Gingold & Monaghan (1977) as an attempt to model continuum physics avoiding the limitations of grid-based finite difference methods. In SPH, fluid elements constituting the system are sampled and represented by particles, and dynamical equations are obtained from the Lagrangian form of the hydrodynamic conservation laws. SPH has two main advantages with respect to other techniques: on one hand, since it follows the evolution of individual fluid elements, the computational resources are put where they are needed most. On the other hand, there is no grid constraining the dynamic range or the global geometry of the systems being studied.

Gravitational interactions between particles can be obtained from different techniques. The most straightforward method consists of computing gravitational accelerations as the direct sum of all interactions between particles. This approach, called particle-particle (PP), has several theoretical advantages over other potential solvers in the context of SPH (Hernquist & Katz 1989). In particular, it is fully Lagrangian, it does not use a grid, and energy conservation is generally better. However, for simulations involving a large number of particles, a unification of SPH and PP techniques was prohibitively expensive because the computing time per step scaled as $\sim N^2$. Gravitational accelerations in SPH codes are then usually computed by using other approaches. For instance, grid-based methods (Monaghan & Lattanzio 1985, Evrard 1988), or the hierarchical tree method (Hernquist & Katz 1989).

Thanks to the development of computers with a data parallel programming model as, for example, the *Connection Machine* (CM), we can now envisage the use of a PP approach in SPH. On this kind of computers, the contribution of a particle to the gravitational accelerations of all the other particles can be calculated in just one parallel operation (Alimi & Scholl 1993, Serna, Alimi & Scholl 1994, Scholl & Alimi 1995). The computing time then scales as $\sim N$ and the practical advantages of other potential solvers do not hold.

We will present in this paper a general-purpose code, called PPASPH, where SPH and PP techniques have been coupled and implemented on CM. After exhaustively testing this code, it has been applied to analyze in detail the following important aspect of the SPH method:

Although SPH allows for an easy implementation of adaptive resolution scales, it introduces additional terms in the equations of motion. These additional terms are usually neglected because they are computationally expensive and, for a high enough number of particles, they are much smaller than the other ones (Gingold & Monaghan 1982, Evrard 1988). However, Hernquist (1993) has found that, when adaptive SPH algorithms are used to simu-late the evolution of adiabatic systems, a simultaneous good conservation of energy and entropy is not obtained. If the thermal energy equation is integrated, the total entropy is not conserved as accurately as the energy. Reciprocally, if an entropy equation is integrated, then the total energy is not conserved as accurately as the entropy. Hernquist (1993) then claimed that conclusions at high resolution using SPH must be accepted with caution. This is in principle also extensible to any other fluid algorithm using adaptive resolution scales or grids. It is then important to analyze in detail if this difficulty can in fact be solved by including the previously neglected terms or, in the opposite, if it reveals some weakness inherent to the SPH technique itself.

This paper is organized as follows. In Section 2 we describe our PPASPH code as well as the implementation of radiative cooling processes. Several tests on this code are then presented in Section 3. The inclusion of additional terms related to the adaptive resolution scales is described in Section 4, as well as the analysis of their importance on the simultaneous conservation of entropy and energy. Section 5 summarizes our main conclusions.

## 2. PPASPH

### 2.1. Basic principles of the SPH method

In SPH, any macroscopic variable (density, pressure gradient,...), $f(\boldsymbol{r})$, is conveniently calculated in terms of its values at a set of disordered points (the particles) by means of an interpolation technique known as kernel estimation. This technique is equivalent to convolving the field $f(\boldsymbol{r})$ with a smoothing, or filter, function $W(\boldsymbol{r} - \boldsymbol{r}', h)$ to produce an estimate of the field, $f_S(\boldsymbol{r})$, where local statistical fluctuations have been smoothed out:

$$f_S(\boldsymbol{r}) = \int f(\boldsymbol{r}')W(\boldsymbol{r} - \boldsymbol{r}', h)d\boldsymbol{r}' , \qquad (1)$$

the integration being over all space. The smoothing length $h$ specifies the extent of the averaging volume and it then determines the local spatial resolution. The smoothing kernel $W(\boldsymbol{r} - \boldsymbol{r}', h)$ is assumed to be spherically symmetric, and normalized to unity when integrated over the space, as part of the requirement $\lim_{h \to 0} f_S(\boldsymbol{r}) = f(\boldsymbol{r})$.

For numerical work, where a finite number $N_g$ of gas particles is used, the integral interpolant is ap-proximated by a summation interpolant:

$$f_S(\mathbf{r}_i) = \sum_{j=1}^{N_g} m_j \frac{f(\mathbf{r}_j)}{\rho(\mathbf{r}_j)} W(\mathbf{r}_i - \mathbf{r}_j, h_{ij}) \;, \qquad (2)$$

where $\rho(\mathbf{r}_j)$ is the density at the position of particle $j$. As a particular case of Eq. (2), the smoothed estimate of the density at $\mathbf{r}_i$ is

$$\rho(\mathbf{r}_i) = \sum m_j W(r_{ij}, h_{ij}) \;. \qquad (3)$$

Here, $r_{ij} = |\mathbf{r}_i - \mathbf{r}_j|$, and $h_{ij}$ denotes a symmetrized smoothing length, $h_{ij} = (h_i + h_j)/2$, necessary to avoid a violation of the reciprocity principle (Evrard 1988). An alternative proposed by Hernquist & Katz (1989) is the use of kernel averages, $[W(r_{ij}, h_i) + W(r_{ij}, h_j)]/2$, instead of an average of smoothing lengths. Both possibilities provide similar accuracies. We have chosen that proposed by Evrard (1988) because it requires a slightly smaller amount of CPU time.

The smoothing formalism also provides a natural means for estimating gradients of the local fluid properties, or any other derivatives. For example, from Eq. (1)

$$[\boldsymbol{\nabla} f(\mathbf{r})]_S = \int \boldsymbol{\nabla}_{\mathbf{r}'} f(\mathbf{r}') W(\mathbf{r} - \mathbf{r}', h) d\mathbf{r}' \qquad (4)$$

or, after integration by parts and approximation by a summation interpolant

$$(\boldsymbol{\nabla} f)_i = \sum_j m_j \frac{f(\mathbf{r}_j)}{\rho(\mathbf{r}_j)} \boldsymbol{\nabla}_i W(r_{ij}, h_{ij}) \;. \qquad (5)$$

## 2.2. Kernels

Many kernel functions can be devised in SPH. That which allows for an easier physical interpretation of any SPH equation is the Gaussian kernel

$$W(r_{ij}, h_{ij}) = \frac{1}{h_{ij}^{\nu} \pi^{\nu/2}} e^{-(r_{ij}/h_{ij})^2} \;, \qquad (6)$$

where $\nu$ is the number of dimensions. Although this kernel interpolates with high accuracy, it has the practical disadvantage that it is not exactly zero for finite $r_{ij}/h_{ij}$ values. Consequently, much particles contribute to local properties and, on sequential and vectorial computers, much computational effort is needed. In order to avoid this inconvenience, several authors use instead kernels based on spline functions with compact support, as that proposed by Monaghan & Lattanzio (1985).

$$W(s_{ij}) = \frac{\sigma}{h_{ij}^{\nu}} \begin{cases} \frac{3}{2}[\frac{1}{2}s_{ij} - 1]s_{ij}^2 + 1 & (0 \leq s_{ij} \leq 1) \\ \frac{1}{4}[2 - s_{ij}]^3 & (1 \leq s_{ij} \leq 2) \\ 0 & (s_{ij} \geq 2) \end{cases} , \qquad (7)$$

where $s_{ij} \equiv r_{ij}/h_{ij}$ and $\sigma$ is a normalization constant with the values $2/3$, $10/7\pi$, $1/\pi$, in one, two, and three dimensions, respectively. This kernel has the practical advantage that it is exactly vanishing for $r_{ij}/h_{ij} > 2$.

In a parallel programming model, the practical advantages of spline kernels do not hold because sums over all the particles can be performed in a parallel way. PPASPH takes then into account all contributions to the local properties, even when they are very small or exactly vanishing. In all the simulations presented in this paper we have used a Gaussian kernel. Nevertheless, other possible choices, as Eq. (7), have been also implemented in PPASPH.

## 2.3. Smoothing Lengths

Ideally, the individual particle smoothing lengths $h_i$ must be updated such that each particle interacts with a constant number of neighbors $N_S$. By *neighbors* we mean those particles $j$ with distances $r_{ij} \leq \mathcal{H}h_i$, where $\mathcal{H}$ is a constant for each kind of kernel[1]. Such a condition can be exactly implemented without additional computing time when gravitational forces are computed using a tree code and neighbor lists are then available. In principle, other gravitational schemes, as the PP one, would need update $h_i$ from conditions like $h_i \propto \rho_i^{-1/3}$. However, this last condition only ensures a roughly constant number of neighbors and some adjustments are needed during a simulation to avoid an excessively large deviation from $N_S$.

We have constructed an efficient algorithm for data parallel programming model which exactly updates $h_i$ while consuming only a modest amount of computing time. Since the identity of neighbors is not necessary in a PP code, we must just find the sphere centered in each particle $i$ which contains a specified number of particles. Such an algorithm has been implemented in the following way.

For each SPH particle $i$, we start with a predictor value $\tilde{h}_i^{n+1} = h_i^n$ of its smoothing length.

---

[1] $\mathcal{H} = 2$ for a spline kernel, while $\mathcal{H} \approx 3$ for a Gaussian kernel

We then count the number of particles $j$ such that $r_{ij} \leq \mathcal{H}\tilde{h}_i^{n+1}$. On CM, this number $N_i$ can be fastly obtained by means of the *COUNT* command. If $N_i$ is not equal to $N_S$, a corrector value is computed from

$$h_i^{n+1} = \tilde{h}_i^{n+1}\frac{1}{2}[1 + (N_S/N_i)^{1/3}] . \tag{8}$$

This corrector value $h_i^{n+1}$ is used as predictor $\tilde{h}_i^{n+1}$ for the next iteration step and the sequence is followed until that $N_i = N_S$. This procedure can be considered as essentially identical to that used by Hernquist & Katz (1989), but with a **zero** tolerance parameter.

However, in order to avoid iterations with an oscillatory convergence, the predictor $h_i$ value given by Eq. (8) is forced to be strictly within the interval $(h_i^1, h_i^2)$, where $h_i^1$ and $h_i^2$ are the last $h$ values found in the iteration sequence such that they imply $N_i < N_S$ and $N_i > N_S$, respectively. Since we iterate until that $N_i = N_S$, our method gives by construction an exactly fixed number of neighbors for all particles. This algorithm is obviously independent of $N_S$ and, on a parallel computer, it just grows with the number of gas particles as $\sim N_g$.

The choice of $N_S$ is determined by the condition that the theoretical density field at initial conditions must be well described by the SPH density field (typically, $N_S \in [30, 50]$). It must be however noted that, if the number of neighbors experiences discrete jumps as $h_i$ for a particle is increased, our zero tolerance algorithm could fail for certain choices of $N_S$. Such a situation is extremely rare in the course of a simulation, but possible at the initial conditions when particles are distributed on a regular lattice. In this case, the input $N_S$ value must be chosen with caution or, otherwise, the number of iterations must be limited to a maximum value.

## 2.4. Dynamical Equations

The evolution of particle $i$ is determined by Euler's equations

$$\frac{d\boldsymbol{r}_i}{dt} = \boldsymbol{v}_i \tag{9a}$$

$$\frac{d\boldsymbol{v}_i}{dt} = -\frac{\boldsymbol{\nabla}P_i}{\rho_i} + \boldsymbol{a}_i^{\text{visc}} - \boldsymbol{\nabla}\Phi_i \tag{9b}$$

where $\Phi_i$ is the gravitational potential at $\boldsymbol{r}_i$, $P_i$ is the local pressure and $\boldsymbol{a}_i^{\text{visc}}$ is an artificial viscosity term allowing for the presence of shock waves in the flow.

The SPH representation of Eqs. (9) is not unique. Several forms of the equations of motion can be found in the literature, none of which appears to be clearly superior to the others. The most often used SPH expression for the pressure gradient and viscosity terms is

$$-\frac{\boldsymbol{\nabla}P_i}{\rho_i} + \boldsymbol{a}_i^{\text{visc}} = \tag{10}$$

$$\sum_{j=1}^{N_g} m_j \left(\frac{P_i}{\rho_i^2} + \frac{P_j}{\rho_j^2} + \Pi_{ij}\right) \boldsymbol{\nabla}_i W(r_{ij}, h_{ij})$$

where $\Pi_{ij}$ stands for the artificial viscosity.

Several expressions for the artificial viscosity have been proposed in the literature. Up to date, that giving the best results in SPH codes is (Monaghan & Gingold 1983):

$$\Pi_{ij} = \frac{-\alpha\mu_{ij}\overline{c}_{ij} + \beta\mu_{ij}^2}{\overline{\rho}_{ij}} , \tag{11}$$

where $\alpha$ and $\beta$ are constant parameters of order unity, $c_i$ is the sound speed at the position of particle $i$, $\overline{c}_{ij} = (c_i + c_j)/2$, $\overline{\rho}_{ij} = (\rho_i + \rho_j)/2$, and

$$\mu_{ij} = \begin{cases} \frac{\boldsymbol{v}_{ij}\boldsymbol{r}_{ij}}{h_{ij}(r_{ij}^2/h_{ij}^2 + \eta^2)} & \boldsymbol{v}_{ij}\boldsymbol{r}_{ij} < 0 \\ 0 & \boldsymbol{v}_{ij}\boldsymbol{r}_{ij} \geq 0 \end{cases} , \tag{12}$$

with $\boldsymbol{v}_{ij} = \boldsymbol{v}_i - \boldsymbol{v}_j$ and $\eta^2$ being a softening constant avoiding numerical divergences. Typically, $\eta^2 = 0.01$.

In order to compute the local pressure, we must specify an equation of state $P_i(\rho_i, u_i)$. For an ideal gas the equation of state is

$$P_i = (\gamma - 1)\rho_i u_i , \tag{13}$$

where $\gamma = 5/3$. Some problems (see section 3.1.2.) could however require the isothermal equation of state $P_i = c_{\text{iso}}^2\rho_i$, where $c_{\text{iso}}$ is the isothermal sound speed. In this last case, an equation for the evolution of the specific internal energy $u_i$ is not needed.

In general, we used the non-symmetrized thermal energy equation:

$$\frac{du_i}{dt} = \sum_{j=1}^{N_g} m_j \left(\frac{P_i}{\rho_i^2} + \frac{\Pi_{ij}}{2}\right) \boldsymbol{v}_{ij}\boldsymbol{\nabla}_i W(r_{ij}, h_{ij}) + \dot{u}^{cool} , \tag{14}$$

which gives better results than a symmetrized equation for the evolution of $u_i$ (Benz 1989).

The term $\dot{u}^{cool}$ appearing in Eq. (14) denotes the cooling (or heating) rates associated with non-adiabatic processes other than shocks.

## 2.5. Implementation of Cooling

The rate of the specific thermal energy variation due to radiative cooling processes is

$$\frac{du_i^{cool}}{dt} = -\frac{\Lambda(T)}{\rho} , \qquad (15)$$

where $\Lambda(T)$ is the cooling function. This function has been obtained by considering the gas as an optically thin 'primordial' mixture of H and He (mass fractions $X = 0.76$, $Y = 0.24$, respectively) in collisional ionization equilibrium at temperature $T_i$. The cooling function is then given by $\Lambda = \Lambda_{brem} + \Lambda_H + \Lambda_{He}$, where $\Lambda_{brem}$ is the bremsstrahlung cooling (Tucker 1975), while $\Lambda_H$ and $\Lambda_{He}$ are the cooling functions for radiative recombination and line emission processes of hydrogen and helium (Bond et al. 1984):

$$\Lambda_{brem} = 7.31 \times 10^{22} f^2 T^{1/2}$$
$$\Lambda_H = 2.32 \times 10^{25} f^2 \frac{T^6}{1 + 0.25T^8} \qquad (16)$$
$$\Lambda_{He} = 7.01 \times 10^{22} f^2 \frac{T^3(1 + 3.6 \times 10^{-5}T^4)}{1 + 3.3 \times 10^{-8}T^8}$$

$f$ being the ionization fraction

$$\frac{f}{1-f} = 3.2 \times 10^4 e^{-13.6/T} T^{1.22} . \qquad (17)$$

Temperatures in Eqs. (16)-(17) are written in eV, while $\Lambda$ is expressed in [erg cm$^3$ g$^{-2}$s$^{-1}$].

Usually, the cooling processes impose strong limits on the integration timestep. These constraints can be however softened by implementing such processes in an integrated form controlled by the Courant condition (see below section 2.8.). As a matter of fact, since this last condition ensures that densities do not change considerably over one step, equation (15) can be integrated to give (Thomas & Couchman 1992):

$$\int_{u_i}^{u_i - \Delta u_i^{cool}} \frac{du_i^{cool}}{\Lambda_i} = -\frac{\Delta t}{\rho_i} . \qquad (18)$$

Then, we have only to find the $\Delta u_i^{cool}$ values which satisfy the above equation.

In some few underresolved zones coincident with the recombination front, the previous method could overestimate cooling processes. An special treatment for these zones must be then included like, for instance, that proposed by Anninos & Norman (1994).

Those authors enforce the pressure equilibrium condition at the cooling front to avoid such a non-physical catastrophic cooling phenomenon.

We have adapted Anninos & Norman's procedure to our SPH code. That is, particles which overcool are selected by using two criteria:

1. The local cooling time ($\Delta t_i^{cool} = u_i/\dot{u}_i^{cool}$) is smaller than the viscous-sound crossing time $\Delta t_{cv}$ (see Eq. [28] bellow):

$$\Delta t_i^{cool} < \Delta t_{cv} . \qquad (19)$$

2. The local pressure after updating the cooling term is smaller than half the average pressure $\overline{P}_i$ of the neighbors of particle $i$

$$P_i \leq \overline{P}_i/2 . \qquad (20)$$

When both criteria are satisfied for a particle, we enforce the condition $P_i = \overline{P}_i$.

## 2.6. Gravitational interactions

Gravitational interactions in PPASPH are computed by using the PP method. This approach has some theoretical advantages over other potential solvers in the context of SPH (see Sect. 1). In addition, it is simpler and easier to parallelize on CM than other methods as, e.g., the hierarchical tree approach.

Since particles in SPH just represent elements of a continuum fluid, gravitational interactions between particles must be smoothed by using the techniques of Sect. 2.1. According to the procedure outlined by Gingold & Monaghan (1977), for a gravitational potential defined as

$$\phi(\boldsymbol{r}) = -G \int \frac{\rho_{tot}(\boldsymbol{r}')d\boldsymbol{r}'}{|\boldsymbol{r} - \boldsymbol{r}'|} , \qquad (21)$$

the smoothed expression of the gravitational acceleration of particle $i$ is

$$\boldsymbol{a}_i = -\sum_{j=1}^{N} Gm_j \left\{ \frac{4\pi}{r_{ij}^3} \int_0^{r_{ij}} W(r, \epsilon) r^2 dr \right\} \boldsymbol{r}_{ij} , \qquad (22)$$

where a gravitational smoothing length $\epsilon$, different from the hydrodynamical one $h$, must be used.

The integral appearing in Eqs.(22) can be analytically solved provided that the functional form of the kernel $W$ has been specified.

For the Gaussian kernel given by Eq. (6), we find

$$\boldsymbol{a}_i = -\sum_{j=1}^{N} \frac{Gm_j}{\epsilon^3} \left\{ \frac{\mathrm{erf}(s_{ij})}{s_{ij}^3} - \frac{2e^{-(s_{ij})^2}}{\pi^{3/2} s_{ij}^2} \right\} \boldsymbol{r}_{ij} \quad (23)$$

$$\phi_i = -\sum_{j=1}^{N} \frac{Gm_j}{\epsilon_{ij}} \frac{\mathrm{erf}(s_{ij})}{s_{ij}} \quad (24)$$

where $s_{ij} \equiv r_{ij}/\epsilon_{ij}$. The corresponding expressions for the spline kernel given in Eq. (7) can be found in the Appendix of Hernquist & Katz (1989).

It must be noted that, when $s_{ij} \gg 1$, the above expressions lead to the Newtonian ones, $\phi_{ij} = -Gm_j/r_{ij}$ and $\boldsymbol{a}_{ij} = -Gm_j \boldsymbol{r}_{ij}/r_{ij}^3$ while, for $s \to 0$, they tend towards a constant value. For a Gaussian kernel we find $\phi_{ij} = -2Gm_j/\epsilon_{ij}\pi^{3/2}$, $\boldsymbol{a}_{ij} = -4Gm_j \boldsymbol{r}_{ij}/3\epsilon_{ij}\pi^{1/2}$. As it has been previously noted by other authors, the SPH technique provides a more coherent way to softening gravitational interactions that the usual $r_{ij} \to (r_{ij}^2 + \epsilon_i^2 + \epsilon_j^2)^{1/2}$ procedure which represents particles as Plummer spheres.

In the above equations $\epsilon_{ij} \equiv (\epsilon_i + \epsilon_j)/2$, where the gravitational smoothing lengths do not change with time, and constitute a minimum value for the variable smoothing length $h$ of the gas particles. The choice of a minimum value for $h$ is similar to setting a maximum density value and, indirectly, a minimum value for the time-step (Navarro & White 1993). The justification for such a condition is that it would be wasteful to estimate the pressure gradients with higher resolution than the gravitational potential in regions where the softening is important.

## 2.7. Time Stepping

The time integration in PPASPH is performed using a PEC (Predict, Evaluate, Correct) variable timestep scheme similar to that considered by Couchman et al. (1994). According to this scheme, one enters the time $t_n$ with known positions $\boldsymbol{r}_n$, velocities $\boldsymbol{v}_n$, and accelerations $\boldsymbol{a}'_n$, for all the $N$ particles, as well as the hydrodynamical quantities (smoothing lengths $h_n$, specific internal energies $u_n$, and their derivatives $\dot{u}_n$) for all the $N_g$ gas particles.

The sequence initiates by predicting variable values (denoted by primes) at $t_{n+1}$ according to

$$\begin{aligned}
\boldsymbol{r}'_{n+1} &= \boldsymbol{r}_n + \boldsymbol{v}_n \Delta t + \boldsymbol{a}'_n (\Delta t)^2/2 \\
\boldsymbol{v}'_{n+1} &= \boldsymbol{v}_n + \boldsymbol{a}'_n \Delta t \\
u'_{n+1} &= u_n + \dot{u}' \Delta t
\end{aligned} \quad (25)$$

The above predicted quantities are then used to compute $\boldsymbol{a}'_{n+1}$ and $\dot{u}'_{n+1}$ at $\boldsymbol{r}'$ by using Eqs. (9b) and (14). These predicted 'forces' are then used to correct the positions, velocities, and thermal energies:

$$\begin{aligned}
\boldsymbol{r}_{n+1} &= \boldsymbol{r}'_{n+1} + \mathcal{A}(\boldsymbol{a}'_{n+1} - \boldsymbol{a}'_n)(\Delta t_n)^2/2 \\
\boldsymbol{v}_{n+1} &= \boldsymbol{v}'_{n+1} + \mathcal{B}(\boldsymbol{a}'_{n+1} - \boldsymbol{a}'_n)\Delta t_n \\
u_{n+1} &= u'_{n+1} + \mathcal{C}(\dot{u}'_{n+1} - \dot{u}'_n)\Delta t_n
\end{aligned} \quad (26)$$

where $\mathcal{B} = 1/2$ is required to obtain accurate velocities to second order, while the choice of $\mathcal{A}$ and $\mathcal{C}$ is somewhat arbitrary. The choice $\mathcal{A} = 0$ gives a scheme similar to a leapfrog scheme except that velocities are predicted forward to the same time as the positions before force evaluation. The choice $\mathcal{A} = 1/3$ nominally gives third-order accuracy for positions (but this is swamped by the error in velocities). We have chosen in this paper $\mathcal{A} = 1/3$ and $\mathcal{C} = 1/2$.

Cooling contributions to the thermal energies are finally included by using the integrated scheme of Sect. 2.5.

The advantages of this PEC scheme as compared to a Runge-Kutta one have been shown by Couchman et al. (1994). They found that, when the timestep is slightly overestimated, SPH simulations using the PEC scheme remain strongly stable while those using a Runge-Kutta one can become exponentially instable.

## 2.8. Timestep length

In order to maintain the numerical integration stability, the timestep must be modified at each step according to different criteria. A first timestep control is that concerning the time scale for significant displacements or changes in velocity due to accelerations:

$$\delta t_a = \min_i \left( \frac{h_i^2}{a_i^2} \right)^{1/4}, \quad (27)$$

where, for dark-matter particles, the smoothing length $h_i$ is replaced by the softening parameter $\epsilon_i$.

A second limit on $\Delta t$ is usually given by a timestep control which combines the Courant and the viscous conditions:

$$\delta t_{cv} = \min_i \left[ \frac{h_i}{c_i + 1.2(\alpha c_i + \beta \max_j |\mu_{ij}|)} \right]. \quad (28)$$

The time integration of shock contributions to the thermal energy equation is also limited by the Courant condition. However, if the cooling function $\Lambda$ is nonzero, an additional timestep control involving $\dot{u}_{cool}$ should be needed and would imply very severe limits on $\Delta t$. Since cooling processes in PPASPH are included in an integrated form with the enforced pressure equilibrium treatment described in Sect. 2.5, such a timestep control is not necessary.

The timestep is then given by

$$\Delta t = \min(\delta t_a, \delta t_{cv}) \ . \tag{29}$$

If the $\Delta t$ resulting from Eq. (29) is greater than some input value $(\Delta t)_{max}$, the timestep is forced to be equal to this upper limit.

## 3. TESTS OF PPASPH

PPASPH has been applied to a number of systems in order to test its ability to reproduce known analytic or numerical solutions.

### 3.1. One-dimensional tests

All 1D-tests presented here were performed with $N = 4096$, $N_S = 40$, $\alpha = \beta = 1$ and $\eta^2 = 0.01$. Dissipational effects, other than those associated with the artificial viscosity, were ignored, as well as gravitational interactions.

#### 3.1.1. The shock tube problem

The one-dimensional shock tube problem proposed by Sod (1978) has become a standard test of all transport and source terms (including artificial viscosities) of hydrodynamic algorithms. It considers a perfect gas distributed on the $x$-axis. A diaphragm at $x_0$ initially separates two regions which have different densities and pressures. All particles are initially at rest. At time $t = 0$ the diaphragm is broken and both regions start to interact. Nonlinear waves are then generated at the discontinuity and propagate into each region: a shock wave which moves from the high to the small pressure region, while the associated rarefaction wave moves in the inverse sense. At the contact discontinuity, the fluid density and specific energy are discontinuous, while the velocity and pressure are continuous. However, in the location of the shock wave, all quantities ($P$, $\rho$, $v$ and $u$) will be discontinuous. The analytical solution to this problem has been given by Hawley et al. (1984), and Rasio & Shapiro (1991).

In our simulation, we have initially considered a $\gamma = 1.4$ perfect gas distributed in the interval $-1 \le x \le 1$ according to:

$$\rho = 1 \quad P = 1 \quad v = 0 \qquad \text{(for } x < 0)$$
$$\rho = 0.25 \quad P = 0.1795 \quad v = 0 \qquad \text{(for } x \ge 0)$$

Fig. 1 shows our results at t=0.15. We see from this figure that our results are in excellent agreement with the analytical solutions. The resulting profiles both in the shock wave (located between $x \sim 0.2$ and $x \simeq 0.25$) and in the contact discontinuity (located at $x \simeq 0.1$) are much less rounded than in previous SPH computations (see, e.g., Monaghan & Gingold 1983, Hernquist & Katz 1989, Rasio & Shapiro 1991) as a result of having used a larger number of particles and, hence, a better resolution.

Much more encouraging for us is the almost complete suppression of postshock oscillations in our results. These oscillations can be seen in the previous SPH simulations of this problem, especially in the velocity field, while no high-frequency vibrations are perceptible in our results. Moreover, the overshoot observed in the velocity field at the tail of the rarefaction wave (located at $x \simeq -0.05$) is smaller in our results than in those previously obtained in the literature.

The weak blip observed in the pressure profile at the contact discontinuity is normal in SPH codes. Such non-physical blip has been explained by Monaghan and Gingold (1983) as due to the fact that the smoothed estimate of pressure is computed by using discontinuous quantities. It is then inevitable that $P$ has some slight perturbation at the contact discontinuity, but it has a negligible effect on the motion.

#### 3.1.2. The isothermal shock problem

Another 1D-problem often used to test hydrodynamics codes is that proposed by Leboeuf, Tajima & Dawson (1979). This problem initially considers an isothermal fluid with a square wave type density profile consisting of a central dense region of density $\rho_h$, surrounded by dilute regions with density $\rho_l < \rho_h$. The fluid evolution is characterized by the rapid formation of a density plateau between rarefaction and shock front regions. The shock front speed, $v_s$, as well as the density $\rho_p$ and the velocity fluid $v_f$ in the plateau region can be analytically obtained as a function of $\rho_h/\rho_l$ by using the solutions given by Leboeuf, Tajima & Dawson (1979).

In our SPH simulation we have considered a initial

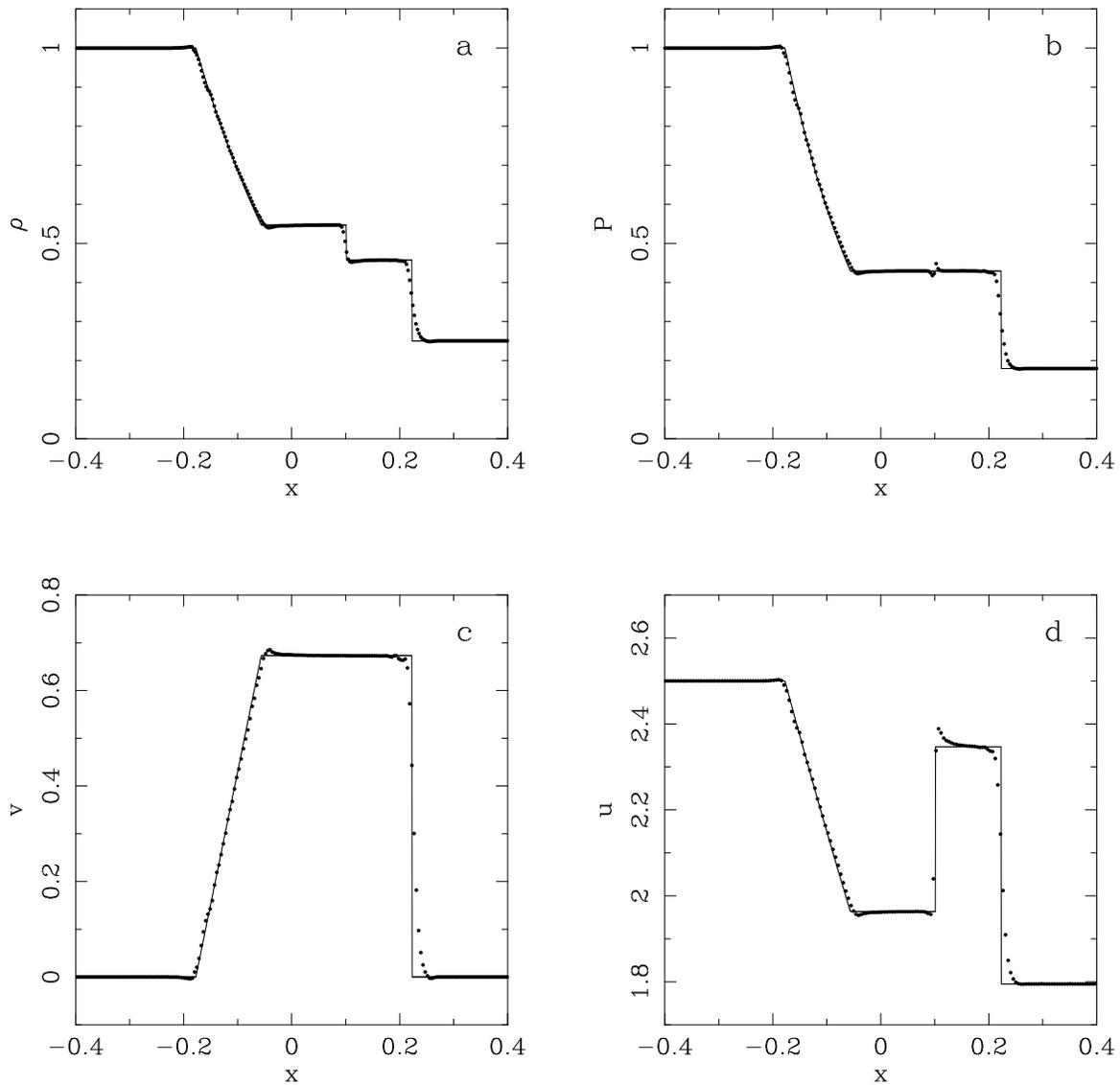

Fig. 1.— a) Density, b) pressure, c)velocity, and d) specific internal energy profiles at t=0.15 in the one-dimensional shock tube problem. Points represent the PPASPH results, and solid lines are the analytical solutions.

density profile given by

$$x < 7 \text{ or } x > 7 \qquad \rho = 4$$
$$-7 < x < 7 \qquad \rho = 9$$

where units were chosen so that $c_s^2 = 1$.

This profile implies the theoretical values $\rho_p = 6$ and $v_f = 0.4$.

The results given by PPASPH are very satisfactory as compared to those previously obtained by other authors (e.g., Gingold & Monaghan 1982, and Haddad, Clausset & Combes 1991) using the SPH method. The theoretical values for $\rho_p$ and $v_f$ are accurately obtained in our simulations (Fig. 2). Moreover, the plateau regions are almost free from high frequency postshock oscillations and broadening effects, which were instead very present in the results reported by the above quoted authors. An overshoot is however observed in the density profile at each corner of the high density region. No other similar overshoots are found in the other discontinuities both in the density and in the velocity profiles.

## 3.2. Adiabatic collapse of a non-rotating gas sphere

A 3D-problem usually considered to test hydrodynamical codes is that concerning the adiabatic collapse of a non-rotating gas sphere. This problem has been studied from a finite-difference method by Thomas (1987), and from SPH simulations by Evrard (1988) and Hernquist & Katz (1989). In order to facilitate the comparison of our results to those obtained by these authors, we have taken their same initial conditions, that is, we initially consider a gas sphere of radius $R$ and total mass $M_T$, with density profile

$$\rho = \frac{M_T}{2\pi R^2} \cdot \frac{1}{r} \qquad (30)$$

All the $N = 4096$ gas particles have initially the same specific internal energy $u = 0.05 G M_T / R$. The ratio of specific heats is $\gamma = 5/3$ and the Gingold & Monaghan artificial viscosity (Eq. 11) was used with $\alpha = 1$, $\beta = 2$ and $\eta^2 = 0.01$. The gravitational softening parameter was $\epsilon = 0.05$ and a Gaussian kernel was assumed with $N_S = 40$. Units were taken so that $G = M_T = R = 1$.

The time evolution of different system profiles are shown in Figs. 3 at times matching those given by Evrard (1988) and Hernquist & Katz (1989). Our results are in excellent agreement with those presented by those authors. Initially far from equilibrium, the system collapses converting most of its kinetic energy into heat (between $t \approx 0.8$ and $t \approx 1.3$). A slow expansion follows and, at late times, a core-halo structure develops with nearly isothermal inner regions and the outer regions cooling adiabatically. The location and strength of the shock are also well reproduced by our code. For example, we find that, at $t = 0.88$, the shock appears located at $r \approx 0.2$ with a Mach number of $\approx 4$. Viscosity erases very efficiently radial motions in the central regions while the collapsing outer regions are rebound and, later ($t \gtrsim 1.3$), they expand constituting a rarefied, adiabatically cooling halo. Internal regions at $t = 0.88$ have $u$-profiles slightly increasing with $r$ but, at the end of our simulation, these regions have a nearly constant temperature. This last feature agrees better with Thomas's finite-difference results than the previously quoted SPH simulations, where a small rise in the thermal energy profile was still present in the final configuration.

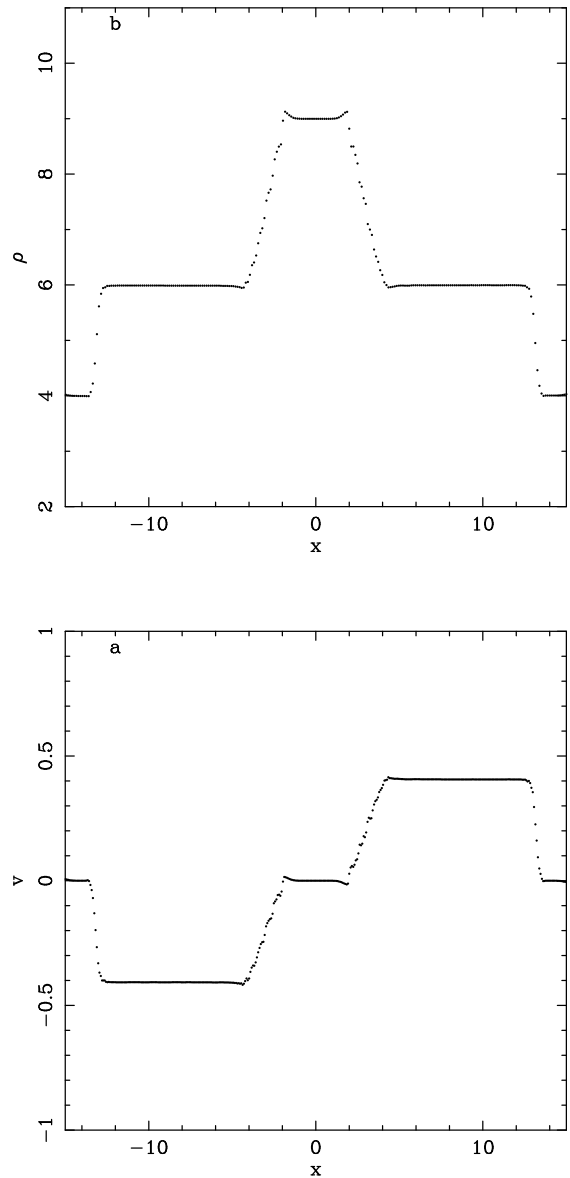

Fig. 2.— a) Density, and b) velocity profiles at t=4 in the one-dimensional isothermal shock problem. Units are defined by $c_s^2 = 1$

Other particular features also agree with those obtained by Evrard (1988) and Hernquist & Katz (1989). For example, the relative maximum observed at $t = 2.2$ in the $v/c$ profile which, although less evident in our simulation, is located at $r \approx 0.6$. Obviously, there exist very small differences between the results reported by all these codes. They just come from the different resolutions and artificial viscosity expressions used in these simulations.

### 3.3. Test of Cooling Simulations

In order to test simulations including radiative cooling processes, we have simulated the collapse of a rotating sphere ('protogalaxy') with a dominant amount of dark matter. This numerical experiment can be compared to that performed by Navarro & White (1993).

Initial conditions consist of $N_{gas} = N_{DM} = 1736$ particles. Positions are obtained from a $\rho(r) \propto r^{-1}$ spherical perturbed grid, while velocities are chosen so that the sphere will be in solid-body rotation around the $z$-axis with a spin parameter $\lambda = J \mid E \mid^{1/2} /GM_{tot}^{5/2} \sim 0.1$ ($J$ and $E$ stand for the total angular momentum and total energy, respectively). The initial radius of the sphere was $R_{tot} = 100$ kpc, and its total mass was $M_{tot} = 10^{12}$ M$_\odot$. The gas represents a 10 per cent of this mass, and is initially at a uniform temperature, $T = 10^3$ K. Gravitational softening parameters were taken to be 2 and 5 kpc for the gas and dark matter, respectively, and units were chosen so that $G = 1$, $[M] = 10^{10}$M$_\odot$, $[L] = 1$ kpc.

When radiative cooling processes are switched-on, the thermal energy gained by particles through shocks is quickly radiated away. Consequently, the gas never gets heated to the virial temperature ($\approx 3 \times 10^6$K). Without pressure support, collapse of the gas proceeds unimpeded until it becomes centrifugally supported in a thin disc-like structure (Fall & Efstathiou 1980, White 1991). As in Navarro & White's (1993) simulation, we find that the disc is almost completely formed few after the collapse time ($t \sim 120$) and evolves little thereafter (see Figs. 4). Shocks dissipate very efficiently the energy in radial motions and the remaining kinetic energy is invested in the rotational motions that support the disc.

A spiral-like structure in the gaseous disc starts to be apparent at $t \approx 160$, and remains during the rest of the simulation. The aspect of this spiral structure at $t = 320$ is intermediate between the two simulations reported by Navarro & White (1993). These authors considered two cases, with gas mass fractions of 0.1 and 0.02, respectively. In the first case, they found a locally unstable disc broken into small clumps while, in the second case, they found a much cleaner spiral structure. Although we used the same gas mass fraction than in the first of such experiences, we find a more locally stable disc. This is certainly due to differences in the initial conditions. The sound speed or the circular frequency are probably higher in our simulation and, consequently, Toomre's (1964) stability parameter is also higher than in the first simulation by Navarro & White (1993). Other features are instead similar to those obtained by the above quoted authors. For example, the presence at $t = 320$ of two or three main spiral arms, and a dense and small core surrounded by a more dilute region.

## 4. CORRECTIONS FOR ADAPTATIVE SMOOTHING LENGTHS

We have previously neglected in Eqs. (9b) and (14) the terms resulting from the fact that $h$ is not a constant. As quoted in the introduction, for not very high numbers of particles, such approximation could introduce non-negligible errors on the SPH results. Hernquist (1993) claimed that the poor entropy conservation observed in some SPH simulations is due to having neglected such terms. Although the global properties of a system do not seem to be altered, ex-cept for a very slight delay in the collapse time of structures, conclusions involving high resolution aspects should be accepted with caution.

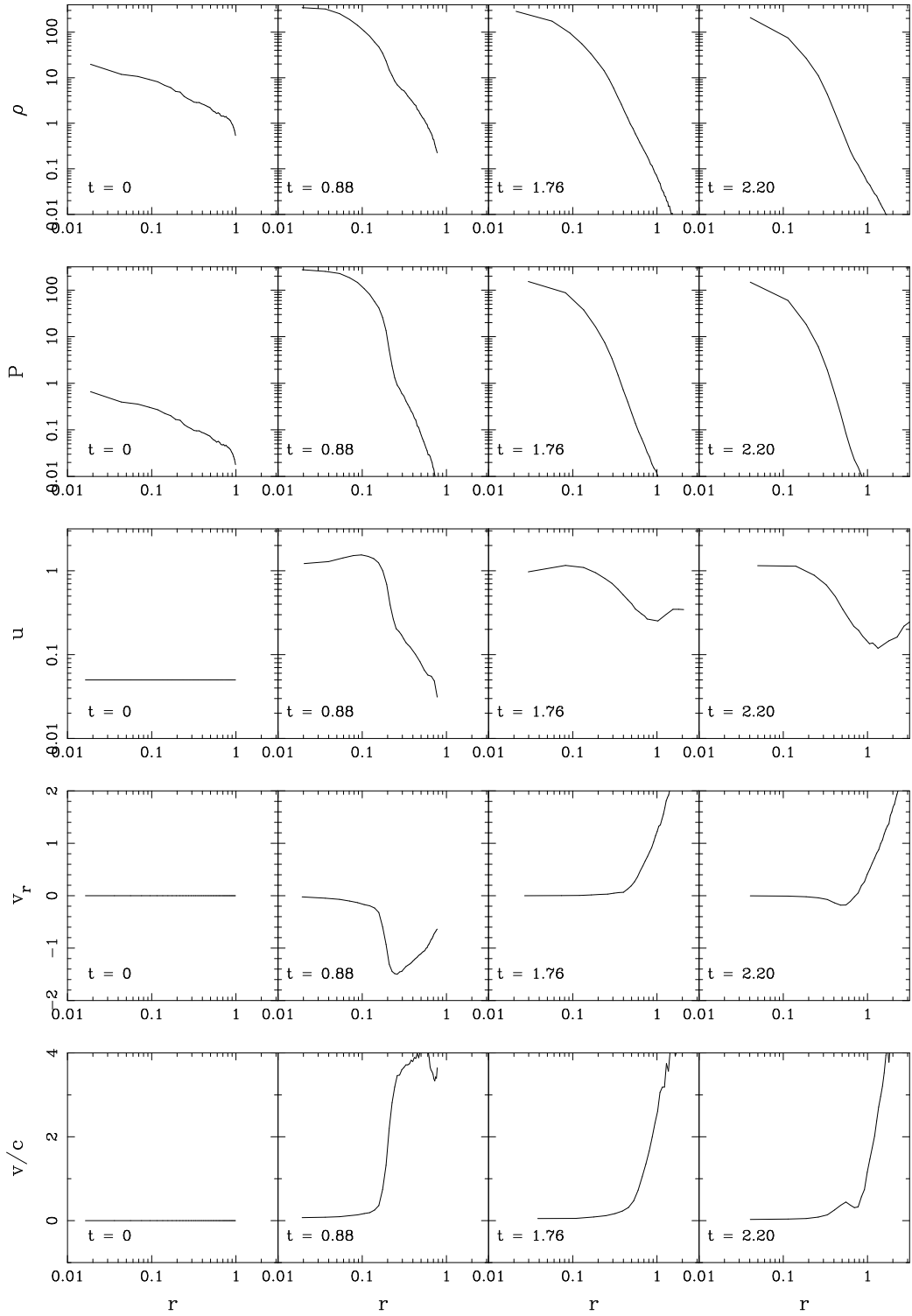

Fig. 3.— Density, pressure, specific internal energy, radial velocity, and Mach number profiles in the adiabatic collapse of a non-rotating gas sphere. The times shown are indicated in the upper right corner of each frame. Units are $G = M_T = R_0 = 1$.

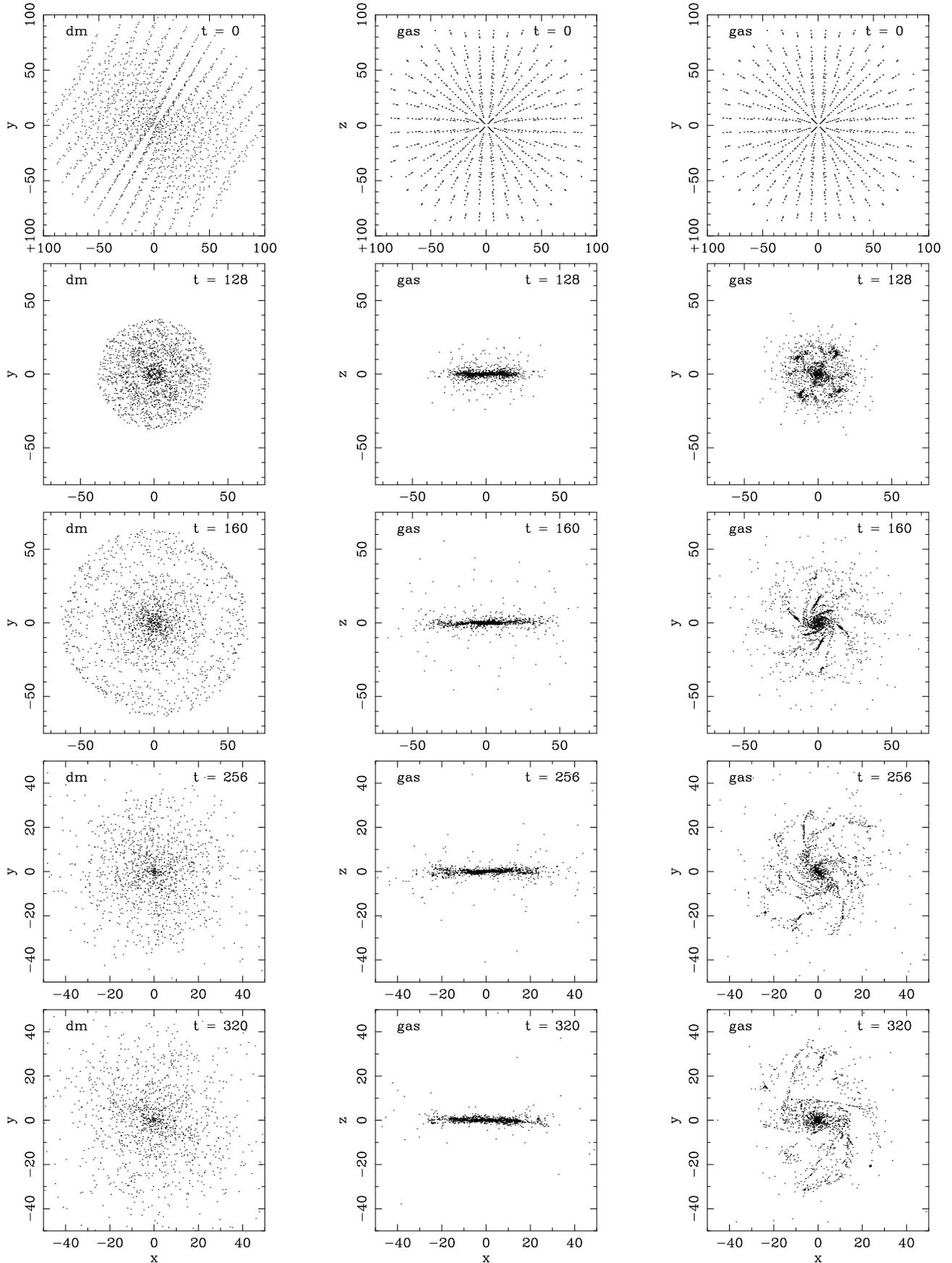

Fig. 4.— Time evolution of a rotating sphere with radiative cooling switched-on. Units are $G = 1$, mass $= 10^{10} M_\odot$, distance $= 1$ kpc

In this section, we will describe how the terms resulting from an adaptative smoothing, called $\boldsymbol{\nabla} h$ terms, have been included in PPASPH, and we will analyze their influence on the entropy conservation.

### 4.1. The $\nabla h$ correction terms

The general expressions for the $\boldsymbol{\nabla} h$ correction terms have been given by Nelson & Papaloizou (1993) in the case of smoothing lengths symmetrized as $(h_i + h_j)/2$ and computed from a procedure of the same type than that described in Sect. 2.3. After some straightforward algebra, Nelson & Papaloizou's expressions can be written in a more compact form as:

$$\tilde{\boldsymbol{a}}_{ij} = -\frac{m_j}{2\mathcal{H}} \left( \frac{P_i}{\rho_i^2} + \frac{P_j}{\rho_j^2} \right) \frac{\partial W_{ij}}{\partial h_{ij}} \frac{\boldsymbol{r}_{ii_m}}{r_{ii_m}} \qquad (31)$$

$$-\delta_{ij_m} \frac{m_j}{2\mathcal{H} m_i} \frac{\boldsymbol{r}_{ij}}{r_{ij}} \sum_{k=1}^{N_g} m_k \left( \frac{P_j}{\rho_j^2} + \frac{P_k}{\rho_k^2} \right) \frac{\partial W_{jk}}{\partial h_{jk}}$$

$$\dot{\tilde{u}}_{ij} = \frac{m_j}{2\mathcal{H}} \frac{P_i}{\rho_i^2} \frac{\partial W_{ij}}{\partial h_{ij}} \left[ \frac{\boldsymbol{v}_{ii_m} \cdot \boldsymbol{r}_{ii_m}}{r_{ii_m}} + \frac{\boldsymbol{r}_{jj_m} \cdot \boldsymbol{v}_{jj_m}}{r_{jj_m}} \right] (32)$$

where $\tilde{\boldsymbol{a}}_{ij}$ and $\dot{\tilde{u}}_{ij}$ are the correction terms to be added in Eqs. (9b) and (14), respectively. Subscripts $k_m$ denote the most distant neighbor of particle $k$, that is, that particle satisfying $| \boldsymbol{r}_k - \boldsymbol{r}_{k_m} | = \mathcal{H} h_k$.

Notice that the above equations require that the most distant neighbor of each particle $k$ be identified. We have implemented it by using the CM function MAXLOC($r_{jk}$, *mask*) which locates the maximum element, $k_m$, of the array containing the distances to $k$ of all particles satisfying the condition *mask* $\equiv r_{jk} \leq \mathcal{H} h_k$. Although this is performed in a parallel way, the function MAXLOC involves communication between processors and, therefore, it breaks somewhat the parallel efficiency of PPASPH. Computing times per step in simulations including $\boldsymbol{\nabla} h$ corrections are typically longer by a factor of two than those neglecting such terms.

### 4.2. Influence of $\nabla h$ correction terms

Simulations as that shown in Sect. 3.2 lead at late times, $t \gtrsim 3$, to an equilibrium sphere with the density profile displayed in Fig. 3. This model is close enough to the collapsed systems which should be found in simulations studying the formation of structures. In order to analyze the influence of the $\boldsymbol{\nabla} h$ correction terms on the entropy conservation, we have simulated the adiabatic evolution of such kind of spheres.

Initial conditions were then generated by performing a simulation like that described in Sect. 3.2, but for different numbers of particles. At $t = 3$, we switched-off its self-gravity and viscous pressures (by setting $\alpha = \beta = 0$) in order to ensure that the subsequent evolution must conserve the total entropy. In absence of gravitational interactions, this system expands fastly and, at $t = 3.3$, its central density has decreased by a factor of $\approx 25$. The evolution from $t = 3$ to $t = 3.3$ must conserve both the total energy, $E$, and the total entropy variable, $S_T = \sum_j m_j \log[a_j(s)]$, where the entropic function $a(s)$ is defined by

$$a(s)_i = \frac{P_i}{\rho_i^\gamma} = \frac{\gamma - 1}{\rho_i^{\gamma-1}} u_i \ . \qquad (33)$$

We have run a series of simulations, where all models started from the same type of initial conditions. Their evolution was always followed by integrating the energy equation (14). The relative variation of energy and entropy from $t = 3$ to $t = 3.3$ is shown in Table 1 (runs labeled by A). We see that, when the $\boldsymbol{\nabla} h$ correction terms are neglected, energy is conserved very accurately but there exists a considerable violation in the total entropy variable (about 5% in the considered time interval). In the opposite, when such correction terms are taken into account, both total energy and total entropy are conserved very accurately (about 0.02%). We thus confirm Hernquist's (1993) interpretation of this entropy violation as a result of having neglected these correction terms.

Inspection of Table 1 shows moreover that the importance of having neglected the $\boldsymbol{\nabla} h$ correction terms does not decrease when a larger number of particles is considered. The entropy violation in these simulations is in fact about 5% whatever the value of $N$ is. However, it must be noted that just taking $N \to \infty$ and $h \to 0$ is not enough to obtain the proper continuum limit. The condition $N_S \to \infty$ is also necessary to ensure that the discrete SPH approach becomes close enough to the continuum physics. Since $h \propto (N_S/N)^{1/3}$, the joint limit $N \to \infty$, $h \to 0$ and $N_S \to \infty$ can be reached by taking $N_S \propto N^\alpha$ with $0 < \alpha < 1$.

In order to analyze more in detail how the lack of entropy conservation depends on $N$ and $N_S$, we have performed some further simulations where $N_S$ was taken as $N_S \propto \sqrt{N}$. In some first test experiences, with initial conditions generated as before, the $\Delta S$ results for different choices of $N_S$ appeared rather influenced by differences in the initial conditions and in the initial small-scale random noise. Error bars in the entropy violation were then rather large ($\Delta S \simeq 5 \pm 2\%$ for the considered time interval) and, hence, it was difficult to extract reliable conclusions. A possible $N_S$ dependence could be masked within the error bars. In order to reduce these initial random fluctuations, we have considered the outcome of the simulation presented in Fig. 3 as that giving the theoretical profiles of a collapsed non-rotating sphere. Particle positions were then settled according to the theoretical density profile by means of the procedure proposed by Whitworth et al. (1995).

The energy and entropy violation found in these numerical experiences is shown in Table 1 (simulations labeled by $B$). We see again that, if the $\boldsymbol{\nabla} h$ correction terms are excluded, the outcoming entropy violation is much larger than that found for the total energy. The lack of entropy conservation does not exhibit any systematic dependence on $N$ or $N_S$. Consequently, if these correction terms are neglected, the SPH results never converge towards those implied by the correct continuum physics, at least in that concerning the entropy.

### TABLE 1
### ENTROPY AND ENERGY CONSERVATION FOR
### DIFFERENT

| Run | $\boldsymbol{\nabla} h$ terms | $N$ | $N_S$ | $\Delta E$ | $\Delta S$ |
|-----|------|------|------|------|------|
| A1 | Excluded | 1024 | 40 | 0.01% | 4.8% |
| A2 | Included | 1024 | 40 | 0.02% | 0.02% |
| A3 | Excluded | 2048 | 40 | 0.01% | 5.1% |
| A4 | Included | 2048 | 40 | 0.02% | 0.01% |
| A5 | Excluded | 4096 | 40 | 0.02% | 5.3% |
| A6 | Included | 4096 | 40 | 0.02% | 0.02% |
| B1 | Excluded | 2048 | 45 | 0.02% | 4.7% |
| B2 | Included | 2048 | 45 | 0.03% | 0.02% |
| B3 | Excluded | 4096 | 64 | 0.02% | 5.0% |
| B4 | Included | 4096 | 64 | 0.02% | 0.03% |
| B5 | Excluded | 8192 | 90 | 0.01% | 4.6% |
| B6 | Included | 8192 | 90 | 0.02% | 0.02% |

Our numerical results can be interpreted as follows: If $\rho_i$ is computed from Eq. (3), its time variation is given by

$$\frac{d\rho_i}{dt} = \sum_j m_j \boldsymbol{v}_{ij} \boldsymbol{\nabla}_i W_{ij} + \dot{\tilde{\rho}}_i \, , \qquad (34)$$

where $\dot{\tilde{\rho}}_i$ denotes the sum of adaptative smoothing corrections terms on $d\rho_i/dt$. Differently from the terms $\dot{\tilde{u}}_i$ and $\tilde{\boldsymbol{a}}_i$, the $\dot{\tilde{\rho}}_i$ term cannot be switched-off when a SPH code as that described in Sect. 2 is used. As a matter of fact, since densities are estimated in practice by evaluating Eq. (3) at each time step, rather than by integrating $d\rho_i/dt$, the $\dot{\tilde{\rho}}_i$ term is always implicitly incorporated.

By differentiating $a(s)_i$ (Eq. [33]) with respect to $t$, we find that the entropic function of particle $i$ changes as

$$\frac{1}{a_i} \frac{da_i}{dt} = \frac{\dot{\tilde{u}}_i}{u_i} - (\gamma - 1) \frac{\dot{\tilde{\rho}}_i}{\rho_i} \, , \qquad (35)$$

where we have replaced $du_i/dt$ and $d\rho_i/dt$ by their expressions (14) and (34).

According to Hernquist's (1993) interpretation, if $\dot{\tilde{u}}_i$ is switched-off, the right hand side of Eq. (35) only contains the term in $\dot{\tilde{\rho}}_i$ and, consequently, any variation of $\rho_i$ should imply a 'non-physical' variation of entropy. In the opposite, if $\tilde{u}_i$ is switched-on, it balances the contribution of $\dot{\tilde{\rho}}_i$ and, in the absence of dissipation, $a(s)$ should be constant particle by particle. The results of our simulations show in fact that both terms balance very accurately. Therefore, $\dot{\tilde{u}}_i/u_i = (\gamma - 1)\dot{\tilde{\rho}}_i/\rho_i$ and, if the adaptative correction terms are switched-off, Eq. (35) becomes

$$\frac{1}{a_i} \frac{da_i}{dt} = -\frac{1}{u_i} \sum_j m_j \frac{P_i}{\rho_i^2} \frac{\partial W_{ij}}{\partial h_{ij}} \frac{\partial h_{ij}}{\partial r_i} \, , \qquad (36)$$

with

$$\frac{\partial h_{ij}}{\partial r_i} = \frac{1}{2\mathcal{H}} \left[ \frac{\boldsymbol{v}_{ii_m} \cdot \boldsymbol{r}_{ii_m}}{r_{ii_m}} + \frac{\boldsymbol{r}_{jj_m} \cdot \boldsymbol{v}_{jj_m}}{r_{jj_m}} \right] \, . \qquad (37)$$

In order to estimate the importance of terms appearing on the right hand side of Eq. (36), we can follow the same kind of simple arguments than those considered by Evrard (1988). That is, we approximate $\partial h_{ij}/\partial r_i \sim h_{ij}/\mathcal{H}$ and, on the other side, we take into account that for a Gaussian kernel:

$$\frac{\partial W_{ij}}{\partial h_{ij}} = -\frac{3}{h_{ij}} \left[ 1 - \frac{2}{3} \left( \frac{r_{ij}}{h_{ij}} \right)^2 \right] W_{ij} \, , \qquad (38)$$

where, since the dominant contribution to the hydrodynamics of any particle comes from scales $r_{ij} \approx h_{ij}$,

we can expect that the average of $[3 - 2(r_{ij}/h_{ij})^2]$ is of order unity. Equation (36) can be then approximated to

$$\frac{1}{a_i}\frac{da_i}{dt} \approx \frac{1}{\mathcal{H}u_i}\frac{P_i}{\rho_i^2}\sum_j m_j W_{ij} = \frac{1}{\mathcal{H}u_i}\frac{P_i}{\rho_i} \, , \qquad (39)$$

and the total entropy variation is nearly given by

$$\frac{dS}{dt} \approx \sum_i \frac{m_i}{\mathcal{H}u_i}\frac{P_i}{\rho_i} \, . \qquad (40)$$

In this equation, the only quantity which depends on the number of particles is $m_i \propto N^{-1}$. The contribution of each particle to $dS/dt$ then decreases as $N^{-1}$. However, since the sum is performed over the $N$ particles, the total entropy variation does not depend on the number of gas particles.

## 5.  SUMMARY

A general-purpose code (PPASPH) for evolving self-gravitating fluids in astrophysics, both with and without a collisionless component has been described.

In PPASPH hydrodynamical properties are computed by using the SPH (Smoothed Particle Hydrodynamics) method, while gravitational forces are computed by a PP (Particle-Particle) approach. A unification of the SPH and PP techniques has several advantages. Since both techniques are gridless, the resulting code is fully Lagrangian and without limitations on the system geometry, or mesh-related limitations on the dynamic range in spatial resolution. The energy conservation is also generally better.

This code has been implemented on the massively parallel computer *The Connection Machine* (CM), which allows for an efficient unification of the SPH and PP methods with costs per time step growing as $\sim N$. Moreover, on CM it is also possible to take into account, with a minimal cost in computing time, the contribution of all particles to the local properties of any other particle. Smoothing lengths can also be updated so that they imply an exactly constant number of neighbors around each particle.

PPASPH has been applied in order to study the importance of correction terms related to adaptative algorithms. We have found that the poor entropy conservation observed in adaptative SPH simulations can in fact be completely improved by taking into account the $\nabla h$ correction terms. In that case, both energy and entropy are conserved with the same degree of accuracy. We thus confirm Hernquist's (1993) interpretation of the entropy violation as a result of having neglected such correction terms. An improvement on the entropy conservation cannot be found by just taking a larger number of particles and/or a larger $N_S$ value. The correct continuum description is only obtained when the $\nabla h$ correction terms are included. Otherwise, the entropy conservation is always rather poor as compared to that found for the total energy.

We thus conclude that SPH conclusions concerning high resolution aspects must be accepted with caution, even if a very high number of particles has been used. For this kind of analyses we believe necessary to perform some additional simulations including the $\nabla h$ correction terms in order to verify that conclusions are not altered. In principle, these cautions ought to apply also to adaptative grid codes. The fact that one can evaluate the correction terms explicitly in SPH constitutes an additional advantage of this approach over grid-based methods.

We thank an anonymous referee for very valuable comments and suggestions. This work was partially supported by the Commisariat à l'Energie Atomique (CEA), Bruyeres le Chatel, France. N-body computations were carried out on the CM-5 of the Centre National de Calcul Parallèle en Sciences de la Terre, Paris, France. A.S. thanks also the Ministerio de Educación y Ciencia, Spain, for a post-doctoral fellowship.

## APPENDIX A.
## IMPLEMENTATION ON CM AND TIMING ANALYSIS

In a simulation using particles, the computationally most expensive parts are those containing a double loop over particles. This is the case in SPH of any smoothed estimate and of gravitational accelerations.

On computers with a data parallel programming model, as the Connection Machine, one physical or virtual processor is assigned to each particle. The contribution of each particle to the smoothed estimate $f$ of all the other particles can be performed by following a direct summation approach with a computing time which grows as $N$. In such approach, $mf/\rho$ is stored as a sequential array $F$ while kernels and distances from particle $j$ to all others are computed and stored as a parallel array $W(:)$. The contribution of $j$ to the smoothed estimate $f$ of all the other particles is then $f_j(:) = F(j) * W(:)$. A single loop over $j$ calculat-ing $f(:) = f(:) + f_j(:)$ will then lead to the $f$ values for all particles. The storage of some quantities both in parallel and in sequential arrays, although not strictly necessary, results in substantial speed gains by avoiding communication between processors. Such storage can be performed very efficiently by using specific CM subroutines.

The above algorithm then computes all interactions disregarding the fact that much of them are almost or exactly vanishing. Other algorithms could be constructed as, for instance, by do not computing contributions when $r_{ij}/h_{ij} > \mathcal{H}$, or by performing the loop over $j$ only for the $N_s$ neighbors of each particle. However, since CM is conceived to perform dummy parallel operations, the first of the above alternatives would need nearly the same time than our algorithm and, in that concerning the second possibility, it would be less efficient on CM because it requires the individual identification of the neighbors of each particle.

TABLE 2
CPU TIMES IN PPASPH (ON CM-5 32 PROC.)

| Section of Code | CPU time | Percentage |
| --- | --- | --- |
| Updating h | 1.85 | 5.6 |
| Hydrodynamics | 4.32 | 13.1 |
| Gravitation | 12.06 | 36.7 |
| Radiative Cooling | 0.56 | 1.7 |
| $\nabla h$ terms | 14.09 | 42.8 |
| Miscellaneous | 0.03 | 0.1 |

For illustrative purposes, we show in Table 2 an example of the CPU distribution in the current version of PPASPH. This example corresponds to the case of the collapse of a $N = 4096$ non-rotating sphere like that considered in Section 3.2, but with radiative cooling processes and $\nabla h$ correction terms switched-on.